\documentclass[10pt, conference, letterpaper]{IEEEtran}

\usepackage{algorithmic}
\usepackage[ruled,vlined]{algorithm2e}
\usepackage{graphicx}
\usepackage{subfigure}
\usepackage{url}
\usepackage{cite}
\usepackage{amssymb,amsthm}
\usepackage{amsmath}
\usepackage{cases}
\usepackage{caption}
\usepackage{amsmath}

\newtheorem{lemma}{\hspace{0.0in}{\bf Lemma}}

\newtheorem{proposition}{\hspace{0.0in}{\bf Proposition}}
\newtheorem{corollary}{\hspace{0.0in}{\bf Corollary}}
\newtheorem{conjecture}{\hspace{0.0in}{\bf Conjecture}}
\def\done{\hspace*{\fill} \rule{1.8mm}{2.5mm} \\ }

\IEEEoverridecommandlockouts

\title{Modeling Dynamics of Online Video Popularity}
\author{
\IEEEauthorblockN{Jiqiang Wu\IEEEauthorrefmark{1}, Yipeng Zhou\IEEEauthorrefmark{2}, Dah Ming Chiu\IEEEauthorrefmark{1}, Youwei Hua\IEEEauthorrefmark{3} and Zirong Zhu\IEEEauthorrefmark{3}}
\IEEEauthorblockA{\IEEEauthorrefmark{1}Department of Information Engineering, The Chinese University of Hong Kong\\}
\IEEEauthorblockA{\IEEEauthorrefmark{2}College of Computer Science and Software Engineer, Shenzhen University\\}
\IEEEauthorblockA{\IEEEauthorrefmark{3}Department of Online Video, Tencent, China\\}
Email: \IEEEauthorrefmark{1}\{wjq010, dmchiu\}@ie.cuhk.edu.hk; \IEEEauthorrefmark{2}ypzhou@szu.edu.cn; \IEEEauthorrefmark{3}\{wwjs, chriszhu\}@tencent.com 
}

\begin{document}
\maketitle

\begin{abstract}
Large Internet video delivery systems serve millions of videos to tens of millions of users on daily basis, via Video-on-Demand (VoD) and live streaming. Video popularity (measured by view count) evolves over time. It represents the workload, as well as business value, of the video to the overall system.  The ability to predict video popularity is very helpful for improving service quality and operating efficiency. Previous studies adopted simple (usually static) models for video popularity, or directly adopted patterns from measurement studies. In this paper, we develop a stochastic fluid model that tries to capture two hidden processes that give rise to different patterns of a given video's popularity evolution: (a) the information spreading process, and (b) the user reaction process. Specifically, these processes model how the video is recommended to the user, the video¡¯s inherent attractiveness, and user¡¯s reaction rate; and yield specific popularity evolution patterns. We then validate our model by matching the predictions of the model with observed patterns from our collaborator, a large content provider in China. This model thus gives us the insight to explain the common and different video popularity evolution patterns and why.
\end{abstract}
\smallskip

\section{Introduction}

Internet VoD streaming and live  streaming are serving a large and increasing number of consumers. Those
large and established content providers maintain a huge catalog of videos and serve tens of millions of users per day.
There is a variety of different video types, for example, movies, TV episodes, news clips, user generated videos, music
videos (MVs), Sports and so on. New videos are added into the system regularly. Some, such as news clips, may have
short lives, whereas others may be viewed over many years. The demand for a given video, over its lifetime (in the
system), is referred to as its popularity (as a function of time). If this demand can be predicted, even roughly, it can be
very helpful for the content provider in running its operation. As can be expected, there are a large number of different
patterns for video popularity evolution, making the job of prediction not so trivial.

Previously, in order to study strategies for system operation, such as content replication and request scheduling,
static workload models for accessing a catalog of videos have been proposed~\cite{rossINFOCOM09,zhouINFOCOM12}. In such static models,
video requests arrive and get served according to stationary stochastic processes, and each request selects a video
according to some static popularity distribution. Such static  models certainly serve a purpose, and
allow us to derive good insights about resource allocation and load balancing. Yet, it is also blatantly clear to us that
in a real-world system, the popularity of different videos are constantly evolving.

A large number of factors can affect video popularity evolution. For example, how is the video recommended?
Depending on its type, a video can be sensitive to its age (measured as the time since it is made available by a given
content provider) to different extents. If it is listed on the web page of the service as one of the top videos (of some sort),
it would tend to draw attention quickly, compared to if it were recommended via social networks. The type of video and its
intrinsic quality or topic can affect the level of potential interest. Given the volume of available catalog of (new) videos,
and relatively constant attention (eyeballs), the rate of user reaction tends to be limited at any given time. The goal of
this study is to arrive at some models that can capture some of the most important underlying factors, and can
represent the way video popularity  evolves in real systems.

There were also previous studies that considered view count as a dynamic value, and tried to come up with
some empirical rules of evolution patterns from analyzing view count logs from real-world systems~\cite{chaTON09,li2012watching,chenICCCN14,pplive13}. This approach,
while it is more based on reality, and quite practical for the given system studied, is nevertheless quite ad hoc and
does not tell us insights that can be applied to different scenarios and systems.

In our study, we go one step further - try to create a model of video popularity evolution based on factors that can be
explained. This is then used to match the real-world view count dynamics of different videos to ensure the model does
represent what is happening in reality. In particular, we try to capture in our model some \emph{hidden process} that is
not easily measurable as the view count dynamics itself.
Before a user decides to view a video, she has to come to know the existence of that video via some form of recommendation,
whether it is via direct marketing, or user-targeted marketing, or word-of-mouth via social networking. In other words, there
is an underlying process to disseminate the information about the video and its availability before users take actions to view
or decide to ignore that video. Once a user becomes aware of the availability of a video, she then makes an independent
decision whether to view it next, based on her interest level and other factors (such as availability of other choices).
We call this the \emph{user reaction process}.
These hidden processes then determine a given video's popularity, and how it evolves over time. These processes can be quite
complicated, depending on a large number of factors. In this paper, we try to keep the model relatively simple by considering
only a few parameters: the rate of direct recommendation, the rate of word-of-mouth recommendation, the video's intrinsic
attractiveness, and the user's reaction rate. We believe by considering these four factors alone, we already have quite a
rich model of evolution dynamics that can match the common evolution patterns observed from our data.

In the rest of the paper, we proceed as follows. First, we describe the data we obtained from our collaborator, a large Internet content
provider in China. We analyze this data and give some observations about common patterns in Sec.~\ref{Sec:measure}.
Then we present our stochastic fluid model of the underlying hidden processes we mentioned above, and derive closed-form solutions
in Sec.~\ref{Sec:Model}. The model and its solutions allow
us to plot the evolution patterns depending on the key parameters of the model. We then go through a case
study of some specific videos and give physical interpretation of their popularity evolution based on our model in Sec.~\ref{Sec:case}.
We also pick out some videos that exhibit some interesting popularity evolution patterns that are not captured well by our model yet. We use these
cases to discuss how our model can be extended in future studies. The related works are discussed in Sec.~\ref{Sec:related}. 
In conclusion, we will discuss the value of our work in creating a new direction for workload and system modeling for large
Internet online video distribution systems, and in applying the results to practical systems.

\section{Measurement}\label{Sec:measure}
We begin our study by conducting a measurement study on a large online video system in China.
This system provides  millions of videos for tens of millions of users per day and the number of active users is more
than 1 million during peak hours. We measure each video's view count based on the
mobile video viewing records collected from the log servers. The dataset covers all viewing records within a six
month period starting from December 1, 2013 to May 31, 2014.
In our study, we analyze the four most important types: movie, music video (MV), News and TV episode.

\begin{figure}[ht]
\centering
\includegraphics[width=0.38\textwidth]{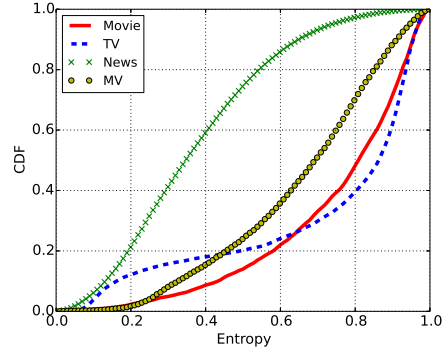}
\caption{CDF of entropy for each video type.}
\label{Fig:entropycdf}
\end{figure}

We study the daily change of view count of a video starting from the day it is uploaded to the online system.
Since each video has a unique set of factors, e.g., recommendation rate, videos are expected to have different view count traces.
Some videos may have a very sharp increase and decrease of popularity; while some videos may have relatively
stable daily view count. To differentiate the view count evolution for different videos, we define a (time window based, normalized)
view count entropy for each video. Let $v_{ji}$ be the number of views of video $j$ on day $i$ since it is uploaded to the online system.
Given total $T$ days in the window, the entropy of video $j$ is
\begin{eqnarray*}
   H_j(T)  & = & -\frac{1}{\ln{T}}\sum_{i=1}^{T}\frac{v_{ji}}{\sum_{i=1}^{T} v_{ji}}\ln{\frac{v_{ji}}{\sum_{i=1}^{T} v_{ji}}}.
\end{eqnarray*}
$\ln{T}$ is the maximum entropy value could be achieved with $T$ days.
Based on the property of entropy, the normalized entropy value should be in between $0 $ and $1$.
If a video's daily view count is stable, its normalized entropy should be close to $1$, otherwise its value will be close to $0$.
We plot the cumulative distribution function (CDF) of 30-days normalized entropy for each type of videos in Fig.~\ref{Fig:entropycdf}.

As we can see, most of the news videos have low entropy value compared to the other three types.
It implies that most news videos have a sharp increase and decrease of popularity in the first 30 days since their birth.
The fractions of videos with low entropy for the other three types are quite small but not zero.
This implies that the change of view count does not purely depend on video type.
In fact, we expect different types of videos are affected by the same factors.
For example, a TV episode recommended through front page may also have a sharp view count change, whereas
an interesting social news not recommended by content providers may have quite stable daily view count.

\begin{figure}[ht]
\centering
\includegraphics[width=0.45\textwidth]{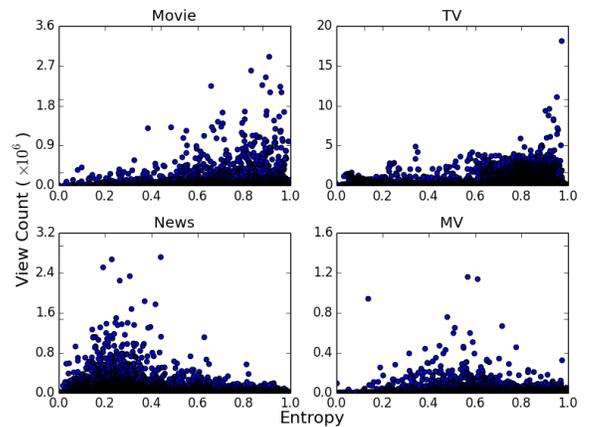}
\caption{Scatter plot of entropy and view count for each type.}
\label{Fig:countventropy}
\end{figure}

To further study the relationship between entropy and view counts, we show the scatter plot in Fig.~\ref{Fig:countventropy},
in which each point represents a video. $x$ axis is the entropy value and $y$ axis is the total view count of 30 days.
Interestingly, we find that the popular movie and TV have large entropy while popular news have small entropy.
We conjecture that the breaking news all got great exposure from front page recommendation
and users' reaction is very quick such that the view count curve of popular news has a  very sharp peak.
For popular movies and TVs, user reaction is not as fast as that of news
because of their long viewing durations, which result in moderate fluctuation of popularity.
MV is a special type with short viewing duration that tends to have fast user reaction but word-of-mouth
recommendation still plays an important role. In Fig.~\ref{Fig:countventropy}, we can see that the pattern of MVs
is quite different from the others. Motivated by the above observations, we propose a model to depict the
evolution of view count in the next section.

\section{Evolution Model}\label{Sec:Model}

Popularity evolution is affected by many factors. It is  too complicated to include all of them.
In this paper, we model four most important factors: rate of direct recommendation, rate of word-of-mouth recommendation,
video attractiveness and user reaction rate.

Different from all previous models about video lifetime and user behavior~\cite{chaTON09,li2012watching,chenICCCN14,pplive13}, we use two processes to describe
popularity evolution. The first process depicts the spread of information about a video; while the second process is the reaction process
once a user has known the video. We believe that these two processes capture what goes on in the real world.

\subsection{Assumptions and Notations}

In our analysis, we make the following assumptions to simplify the problem:
\begin{itemize}
\item The total user population is fixed. In practice, user population at a content provider is not
fixed. A video's views will be influenced by
the change of user population. The model of popularity  evolution does not consider
user population churn.
\item Videos are independent, i.e., each user independently selects videos. In reality, a user who have watched
or dislike a video may refuse to view other related videos, which means user selection is not independent.
We believe the chance of such events is low so that they are ignored in our model.
\item Videos are not replayed by the same users.  In practice, users may replay a good video from time to time.
However, for most videos, previous study showed that it is rarely replayed by the same user~\cite{chenICCCN14}.
\end{itemize}
Without loss of generality, we use video $j$ to describe our model.
The notations used to create the evolution model are listed in Table~\ref{Tab:Notation}.
\begin{table}[ht]
\centering
\caption{List of Notations}
\label{Tab:Notation}
\begin{tabular}{|l|l|}
  \hline
  $N$ & total number of users \\
  $\mathcal{X}_t$ & at time $t$, the set of users who know and would watch video $j$\\
  $x(t)$ & the number of users  in $ \mathcal{X}_t$\\
  $\mathcal{Y}_t$ & at time $t$, the set of  users who know but refuse to watch video $j$ \\
  $y(t)$ &the  number of users  in  $\mathcal{Y}_t$\\
  $\mathcal{S}(t)$ & at time $t$, the set of users who do not know video $j$ \\
  $s(t)$ & the number of users in set  $\mathcal{S}(t)$\\
  $\mathcal{Z}_t$ & at time $t$, the set of users who know video $j$ but do not react yet \\
  $z(t)$ &the  number of users in set $\mathcal{Z}_t$ \\
  $\mathcal{W}_t$ & the set of users who have viewed video $j$ on or before time $t$.\\
  $w(t)$ & the number of  users in set  $\mathcal{W}_t$ \\
  $\alpha$ & rate of direct recommendation  \\
  $\beta$ & rate of word-of-mouth recommendation \\
  $q$ & video attractiveness \\
  $\gamma$ & users' reaction rate \\
  \hline
\end{tabular}
\end{table}

Basically speaking, there are two kinds of notations. The first kind represents various user populations,
classified based on their knowledge or action. The second kind consists of those parameters associated with each video,
determined by both intrinsic and extrinsic factors. These parameters must
be greater than $0$. For example, $\gamma$, the rate that users in $\mathcal{Z}_t$ view video $j$, must be greater than $0$.
$\gamma$ may depend on many factors such as the video $j$'s type, duration, video content and so on.

\subsection{Model of Information Spreading}

The first process is the spread of information about the video. Direct recommendation and word-of-mouth recommendation
are two main methods of distributing a video's information.
This process can be modeled through the following ordinary differential equations (ODEs).
\begin{eqnarray}
 \label{EQ:dxt}
   \frac{dx(t)}{dt} &=& \Big(\alpha s(t) + \beta s(t) x(t)\Big)q \\
   \label{EQ:dyt}
   \frac{dy(t)}{dt} &= &\Big(\alpha s(t) + \beta s(t) x(t)\Big)(1-q) \\
   \label{EQ:dst}
   \frac{ds(t)}{dt} &= &-\frac{dx(t)}{dt} - \frac{dy(t)}{dt}
\end{eqnarray}
$x(t)$ is the number of users who know video $j$ and have interest in viewing video $j$
sooner or later; while $s(t)$ is the number of users who do not know video $j$.
Through direct recommendation, users in $\mathcal{S}_t$ get knowledge of video $j$ with constant rate $\alpha$.
Through word-of-mouth recommendation, users leave  $\mathcal{S}_t$ with rate $\beta x(t)$.
$q$, video $j$'s attractiveness,  determines the final user population
who will view video $j$. The initial conditions are $x(0)=0$, $y(0)=0$ and $s(0)=N$.
Given the fixed total user population, we have
\begin{eqnarray}
 \label{EQ:totalN}
N& =& x(t) + y(t) + s(t).
\end{eqnarray}
By substituting Eq.~\ref{EQ:totalN} to Eq.~\ref{EQ:dxt}-\ref{EQ:dst}, we can get
\begin{eqnarray}
\label{EQ:xt}
   x(t) &=& \frac{qNg(t)-\frac{\alpha}{\beta}}{g(t)+1} \\
      \label{EQ:yt}
   y(t) &=& \frac{1-q}{q}x(t) \\
\label{EQ:st}
  s(t)  &=& N-\frac{x(t)}{q}
\end{eqnarray}
Here $g(t) = e^{(\alpha+\beta q N)t+ \ln \frac{\alpha}{\beta q N}}$.
The detailed derivation is in the appendix.

\begin{lemma}
\label{LEM:Limitxys}
As time $t$ approaches infinity, $x(t)$ approaches $Nq$ from below, $y(t)$ approaches $(1-q)N$ from below and
$s(t)$ approaches $0$ from above.
\end{lemma}
The proof is straightforward since each user in $\mathcal{S}_t$ either joins $\mathcal{X}_t$ with probability $q$ or
joins $\mathcal{Y}_t$ with probability $1-q$. As a result, $s(t)$ keeps decreasing and approaches $0$.

Let $\tau =\alpha+\beta qN$, by differentiating $g(t)$, we have
\begin{eqnarray*}
   g'(t) & = &\tau g(t)>0 ,\\
   g''(t)& = &\tau^2 g(t)>0 ,\\
   g^{(3)}(t) & = & \tau^3 g(t)>0.
\end{eqnarray*}
Since, only the users in $\mathcal{X}_t$ will watch the  video sooner or later,
we focus on analyzing the evolution of $x(t)$ and $x'(t)$.
Based on differentiations of $g(t)$ together with Eq.~\ref{EQ:xt}, we get
\begin{eqnarray*}
\label{EQ:diffxt}
   \frac{dx(t)}{dt} &=&\frac{\tau^2g(t)}{\beta(g(t)+1)^2}>0, \\
\frac{d^2x(t)}{dt^2} &=& \frac{\tau^3g(t)(1-g(t))}{\beta(g(t)+1)^3},\\
  \frac{d^3x(t)}{dt^3} &=& \frac{\tau^4g(t)(g^2(t)-4g(t)+1)}{\beta(g(t)+1)^4}.
\end{eqnarray*}

Letting $\frac{d^2x(t)}{dt^2} = 0$, we have $t'= \frac{1}{\alpha+\beta qN}\ln\frac{\beta q N}{\alpha}$.
If $\alpha> \beta qN$, $t'<0$; while if $\alpha<\beta q N$, $t'>0$. With different $\alpha$, we expect to
have different curves for $x(t)$. Letting $x^{(3)} = 0$, we get two roots:
\begin{eqnarray*}
t_1 =  \frac{1}{\alpha+\beta qN}\ln{\frac{(2+\sqrt{3})\beta q N}{\alpha}},\\
t_2 =  \frac{1}{\alpha+\beta qN}\ln{\frac{(2-\sqrt{3})\beta q N}{\alpha}}.
\end{eqnarray*}
It is trivial to show that $t_2<t'<t_1$. Now, we discuss the evolution of $x(t)$ and $x'(t)$ in two cases.

\subsubsection{ Case I: $\alpha>\beta qN$}

\begin{figure}[ht]
\centering
\includegraphics[width=0.38\textwidth]{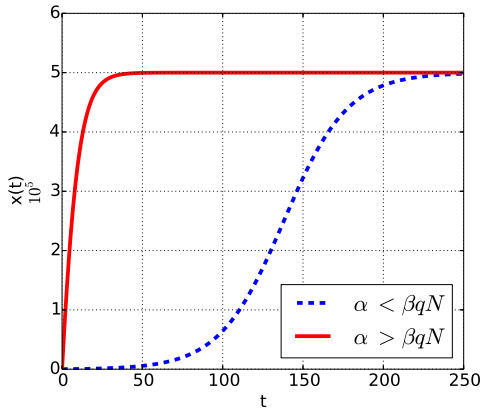}
\caption{ Evolution of $x(t)$ with $N=1,000,000$, $\beta=0.1/N$, $q=0.5$, $\alpha=0.00005 \text{ or }  0.1$. }
\label{Fig:numxt}
\end{figure}

\begin{proposition}
\label{Prop:xtcurve1}
If $\alpha> \beta qN$, $x(t)$ increases as a concave function with time $t$ .
\end{proposition}
The proof is straightforward. Because $\alpha> \beta qN$, $t'<0$. $x''(t)<0$  for $t>0$.
Meanwhile $x'(t)>0$, thus $x(t)$ increases as a concave function with time $t$.
The curve of $x(t)$ in this case is shown with solid line in Fig.~\ref{Fig:numxt}.

\begin{proposition}
\label{Prop:dxtcurve1}
If $\alpha> \beta qN$, $x'(t)$ is a monotonically decreasing function with time $t$.
If $\beta qN <\alpha< (2+\sqrt{3})\beta qN$, $x'(t)$ has two stages:
decreases as a concave function in $\left[0, t_1\right]$;
decreases as a convex function in $\left[t_1, +\infty\right)$.
If $ \alpha> (2+\sqrt{3})\beta qN$, $x'(t)$ decreases as a convex function in $\left[0, +\infty\right)$.
\end{proposition}
The detailed proof is in the appendix. In Fig.~\ref{Fig:numdx}(b), we plot the curves of $x'(t)$ for the case $\alpha > \beta qN$. 

\begin{figure}[ht]
\centering
\includegraphics[width=0.48\textwidth]{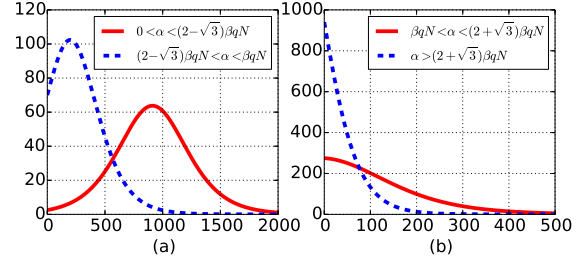}
\caption{ Evolution of $dx(t)/dt$ with  $N=1,000,000$, $\beta=0.1/N$, $q=0.05$, $\alpha = 0.00005 \text{ or } 0.0014 \text{ or } 0.0055\text{ or } 0.0188$.}
\label{Fig:numdx}
\end{figure}

\subsubsection{Case II: $\alpha<\beta qN$}
\begin{proposition}
\label{Prop:xtcurve2}
If $\alpha<\beta qN$, $x(t)$ is a ``S'' curve. It increases as a convex function from $t=0$ to
$t=t'$ and increases as a concave function since $t>t'$.
\end{proposition}
The  proof is straightforward. If $\alpha<\beta qN$, then $t'>0$. In $[0,t']$, $x''(t)>0$, which implies that $x(t)$
increases as a convex function; in $[t', +\infty)$, $x''(t)<0$, which implies that $x(t)$ increases as a concave function.
The curve of $x(t)$ in this case is illustrated with dashed line in Fig.~\ref{Fig:numxt}.

From Proposition~\ref{Prop:xtcurve1} and Proposition~\ref{Prop:xtcurve2}, we can find that direct recommendation and word-of-mouth recommendation 
have different impact on the spread of video information. If direct recommendation dominates information spread, i.e., $\alpha>\beta qN$,
$x(t)$ is a concave curve and $dx(t)$ is a monotonically decreasing curve, otherwise $x(t)$ is a ``S'' curve and the curve of
$dx(t)$ is an inverse ``V''.

\begin{proposition}
\label{Prop:dxtcurve2}
If $\alpha< \beta qN$, $x'(t)$ is a monotonically increasing function in $[0,t']$ and a monotonically decreasing
function in $[t', +\infty)$. If $\alpha<(2-\sqrt{3})\beta qN$, $x'(t)$ has four stages:
increases as a convex function in $[0, t_2]$; increases as a concave function in $[t_2, t']$; decreases as a concave function in $[t', t_1]$;
decreases as a convex function in $[t_1, +\infty)$. If $(2-\sqrt{3})\beta qN<\alpha <\beta qN$, $x'(t)$ has three stages: increases as a concave function in $[0, t']$; decreases as a concave function in $[t', t_1]$; decreases as a convex function in $[t_1, +\infty)$.
\end{proposition}
The detailed proof is in the appendix. Fig.~\ref{Fig:numdx}(a) illustrates the curves of $x'(t)$ for the case $\alpha < \beta qN$.

\begin{corollary}
$x'(t)$ has a unique maximum point, achieving either at $t=0$ if $\alpha >\beta qN$ or at $t=t'$
if $\alpha<\beta qN$.
\end{corollary}
Intuitively speaking, $\alpha$ as the direct recommendation rate and $\beta$ as the word-of-mouth recommendation rate have
different impacts on $x'(t)$. If $\alpha $ dominates information spreading, $x'(0)$ is the maximum value; otherwise
$x'(t')$ is the maximum value. $x'(t)$ will significantly affect view count. We conjecture that if $\alpha$ dominates,
view count evolution has light tail; otherwise view count evolution has heavy tail.

\subsection{Model of User Reaction}

The second process depicts how users react after they know video $j$.
$\gamma$ is the rate to watch video $j$ by those users in $\mathcal{X}_t$ who have not viewed video $j$ yet.
The second process can be captured by the following two ODEs.
\begin{eqnarray}
\label{EQ:dzt}
   \frac{dz(t)}{dt} &=& -\gamma z(t) + \frac{dx(t)}{dt} \\
   \label{EQ:dwt}
   \frac{dw(t)}{dt} &=& \gamma z(t)
\end{eqnarray}
$\mathcal{Z}_t$ contains those users in $\mathcal{X}_t$ who have not viewed video $j$ yet by time $t$.
Naturally, $z(t)$ increases with rate $\frac{dx(t)}{dt}$ and decreases  with rate $\gamma z(t)$.
$w(t)$ is the number of users who have viewed video $j$ up to time $t$. $\frac{dw(t)}{dt}$ is the
number of views of video $j$ at time $t$ and can be used to calculate video popularity at time $t$. 

By solving Eq.~\ref{EQ:dzt}, we have
\begin{eqnarray}
\label{EQ:zt}
   z(t) & = & e^{-\gamma t} \int_{t_0=0}^{t_0=t} x'(t_0)e^{\gamma t_0}dt_0.
\end{eqnarray}
The derivation of Eq.~\ref{EQ:zt} is in the appendix. With Eq.~\ref{EQ:zt}, we can numerically solve $z(t)$ and $\frac{dw(t)}{dt}$.
It is quite difficult to get the closed-form solution of $z(t)$ from Eq.~\ref{EQ:zt}.
By splitting time into discrete time slots, we can still prove a useful proposition.
\begin{proposition}
\label{Prop:dwt}
The curve of $\frac{dw(t)}{dt}$ has a unique peak value. It occurs at the time no earlier than
the time when $x'(t)$ achieves maximum value. In other words, a video's view count either decreases monotonically or increases
monotonically before it decreases monotonically.
\end{proposition}
The detailed proof is in the appendix. Intuitively speaking, the peak of video views occurs after the peak of $x'(t)$.

\begin{conjecture}
\label{Conj:peakdw}
The peak of $\frac{dw(t)}{dt}$ is affected by $\gamma$. With larger $\gamma$, i.e., faster users reaction,
the peak time of $\frac{dw(t)}{dt}$ is closer to the peak time of $x'(t)$.
\end{conjecture}
Without closed-form solution of $z(t)$ or $\frac{dw(t)}{dt}$, it is difficult to prove this conjecture rigorously.
We explain and justify this conjecture through plotting numerical solutions.

\begin{corollary}
\label{Coro:biggamma}
As $\gamma $ approaches infinity, $\frac{dx(t)}{dt} \approx \frac{dw(t)}{dt}$.
\end{corollary}
Larger $\gamma$ implies that users' reaction is fasters when they know the video.
In the extreme case as $\gamma $ approaches infinity, all users will play the video once they know the
availability of the video. In this case, the curve of $\frac{dw(t)}{dt}$ is almost the same as that of $x'(t)$.

\begin{figure}[ht]
\centering
\includegraphics[width=0.48\textwidth]{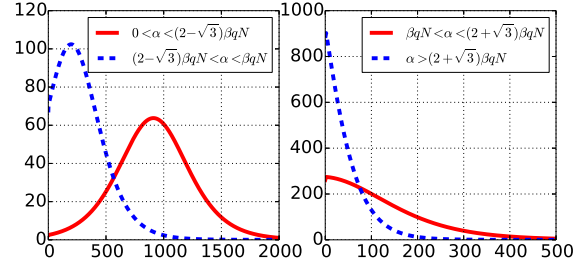}
\caption{Evolution of $dw(t)/dt$ with $\gamma=10$ and other parameters the same with $dx(t)/dt$ in Fig.~\ref{Fig:numdx}.}
\label{Fig:numdwgh}
\end{figure}

\begin{figure}[ht]
\centering
\includegraphics[width=0.48\textwidth]{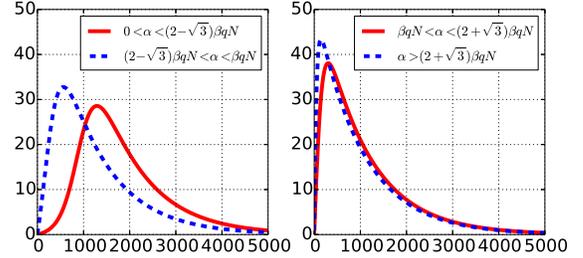}
\caption{Evolution of $dw(t)/dt$ with $\gamma  = 0.001$ and other parameters the same with $dx(t)/dt$ in Fig.~\ref{Fig:numdx}.}
\label{Fig:numdwgl}
\end{figure}

Fig.~\ref{Fig:numdwgh} and Fig.~\ref{Fig:numdwgl} show the curves of $\frac{dw(t)}{dt}$ with different $\gamma$. In Fig.~\ref{Fig:numdwgh}, we set $\gamma=10$ a high value.
As one can observe, the evolution of $\frac{dw(t)}{dt}$ is almost the same as the one
of $\frac{dx(t)}{dt}$ in Fig.~\ref{Fig:numdx},  which has been stated in Corollary~\ref{Coro:biggamma}.
In Fig.~\ref{Fig:numdwgl}, we set $\gamma = 0.001$ to get different evolution of $\frac{dw(t)}{dt}$.
Compared with Fig.~\ref{Fig:numdwgh},
for smaller $\gamma$ the peaks of $\frac{dw(t)}{dt}$ are postponed.

\section{Case Study}
\label{Sec:case}

One way to validate our analytical model of video popularity (view count) evolution is via an independent simulation model. This, however, does not exist and even if we build one, it would involve various additional assumptions and parameters. Instead, we turn to the measurement data collected from our collaborator's real world online video distribution system, and use that to give credibility to our analytical results. More specifically, we will use typical videos' view count history as case studies, and match their patterns to what our model predict using specific parameters. Based on the meaning of these parameters and other information about these videos known to us, we explain why the model matches with the actual behavior.

Two typical videos are selected for each type. \emph{Sample 1}
is a popular video with high attractiveness $q$ and is usually recommended through front page by content providers.
\emph{Sample 2} is an unpopular video with much less total views.
In addition to the discussion of normal cases that can be explained by our model, we introduce
two more cases that are not modeled well. The first one is affected by constrained eyeball while the second one is affected by interest shift. The two additional
cases indicate that the quantified factors, e.g., $\alpha$ and $q$, are not always constant and
extending our model to the cover such cases will be our future work.

In the case study,  we plot the evolution of normalized daily view count.
Take video $j$ as an example, if the peak daily view is $v_{j}^{*}$, then the normalized view count of the $t^{th}$ day
is $\frac{v_{j}(t)}{v_{j}^{*}}$. This way, we can compare the evolution of view counts of two videos
with very different peak daily view counts. Note, we should compare the popularity evolution with the same
number of days. However, videos are uploaded at different times such that the number of
days observed for each video is different. We will try to compare videos with about the same
number of observed days, though it cannot be strictly realized.

\subsection{Normal Case Study}

\begin{figure}[ht]
\centering
\includegraphics[width=0.38\textwidth]{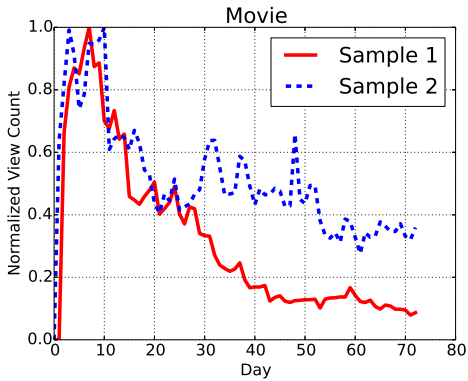}
\caption{Case study of movie. }
\label{Fig:csmovie}
\end{figure}

For movies, we list two patterns of the  popularity evolution.  The first pattern includes most popular
videos with high attractiveness and high view count, which are usually recommended by content providers and favored by most users.
The second pattern includes unpopular videos with low attractiveness and low total view count.
We select two movies, both recommended by the content provider\footnote{
Most new movies are recommended by content providers, since the update speed of movies is very slow. }.
Sample 1 is an American movie called ``Now You See Me'', directed by  Louis Leterrier. This video's
attractiveness is very high. The rating given by those users who have viewed this video is $9.2$ (over max rating $10$).
It is played for about 8 million times since it is uploaded. During our observation period,
the peak view is about $0.1$ million.  Sample 2 is a Chinese movie called ``Perfect Beyond'' , also recommended
by the content provider. There are no famous Chinese stars in this movie, and thus its attractiveness is low with no more than $0.4$ million
total views. However, the rating of ``Perfect Beyond'' is $7.1$, implying some users will spread information about this movie
by word-of-mouth recommendation.  We plot their normalized daily view counts in Fig.~\ref{Fig:csmovie}.

$x$ axis is the number of days since the video is uploaded and $y$ axis is the normalized view count. As
we can see, the curve of sample 1 has a sharper peak than that of sample 2. Since word-of-mouth recommendation plays a more important
role, the tail of sample 2's evolution is heavier.

\begin{figure}[ht]
\centering
\includegraphics[width=0.38\textwidth]{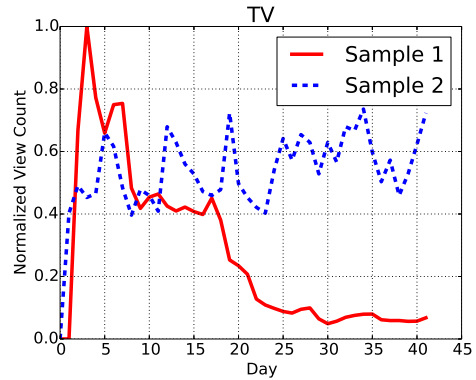}
\caption{Case study of TV. }
\label{Fig:cstv}
\end{figure}

For TV, we also give two evolution patterns as case study. The first pattern includes popular TVs, usually
recommended by content providers, is expected to have popularity evolution with a sharper peak. The second pattern includes
unpopular TVs, usually not recommended by content providers. Sample 1 video is the most popular TV episode in 2014 in the collected data. It is the first episode of ``Perfect Couple'', which is played more than $18$ million times within the first 40 days.
Sample 2 is the last episode of TV ``Mysterious Transfer Student''. It is an unpopular TV episode with only $40$ thousand total views.  This TV
is produced by Japan. Only a small fraction of Chinese viewers are interested in it. We plot the evolution of view counts
for both episodes in Fig.~\ref{Fig:cstv}. $x$ axis and $y$ axis have the same meaning with Fig.~\ref{Fig:csmovie}.
As we can see, since the information spreading rate by direct recommendation of the most popular
TV is very high, its popularity evolution is very sharp. The evolution of the second one, mainly replying on word-of-mouth recommendation
to spread information, is rather stable.

\begin{figure}[ht]
\centering
\includegraphics[width=0.38\textwidth]{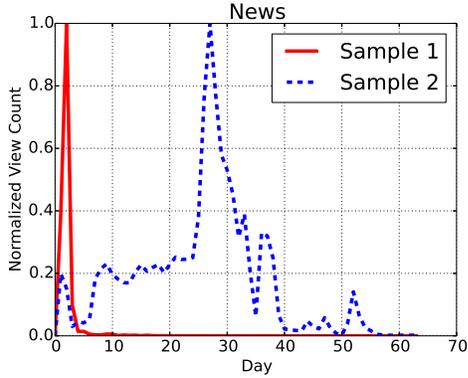}
\caption{Case study of news. }
\label{Fig:csnews}
\end{figure}

News is a special content type since most news lose their attractiveness to users very quickly with time.
We consider two kinds of news videos with two evolution patterns. The first pattern is for breaking news and is expected to have a
large direct recommendation rate and very fast user reaction rate. Popularity evolution should be very sharp and
the peak should appear quite early. The second pattern is not very sensitive to time, e.g., popularity evolution of social news.
It is expected that the second pattern should have a stable evolution. Mainly relying on word-of-mouth recommendation to spread information,
the peak time should be later than the first pattern. In this study, sample 1 is the most popular breaking news in 2014 about
the missing of MH370. It achieves roughly $80$ million of total views within one week time. Sample 2 is a social news about uncivilized behavior
in Beijing subway. Normalized view counts of the two samples are plotted in Fig.~\ref{Fig:csnews}. $x$ axis and $y$ axis
have the same meaning with Fig.~\ref{Fig:cstv}. From the popularity evolution, we can find that almost all views of sample 1 occurred
in the first few days. Sample 2, replying on word-of-mouth recommendation to spread information, has a peak at the 27th day since its birth.

\begin{figure}[ht]
\centering
\includegraphics[width=0.38\textwidth]{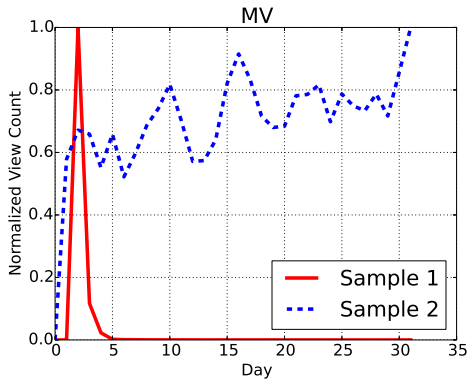}
\caption{Case study of music video. }
\label{Fig:csmv}
\end{figure}

For music videos, we also find two types of videos for case study. The first type is about the latest moves of some
popular star. These videos, usually recommended by content providers, can attract fans' attention in a very short period.
Thus, they are expected to have a very sharp peak in popularity evolution.
The second type is the music created by singers themselves. User reaction rate is usually not very fast and information spreading relies on word-of-mouth recommendation. These videos are expected to have a stable popularity evolution.
In this study, sample 1 is a song sung by famous Hong Kong movie star Maggie Cheung Man-yuk in Shanghai. As a movie star,
Maggie Cheung is not as good as a singer. Her MV attracted a lot of fans but was not well received.
It was played about $3$ million times within one week time.
Sample 2 is a good music video with very high user rating $9.2$. However, mainly relying on word-of-mouth recommendation for information spreading,
its total view count is not high during our measured time window and its popularity evolution is quite stable. We plot two samples in Fig.~\ref{Fig:csmv}, which is consistent with our analysis.

\subsection{Cases not modeled well}

We mainly model four factors: direct recommendation, word-of-mouth recommendation, video attractiveness and user reaction.
For most cases, we can give a good explanation of the popularity evolution if videos' view count is mainly
affected by these four factors and these factors are unchanged with time. However, from our data, we also found some interesting cases that are not modeled well by our existing model. We mainly show two cases, and explain why their behavior deviated from our model. They serve as good leads for us to refine our model.

\begin{figure}[ht]
\centering
\includegraphics[width=0.38\textwidth]{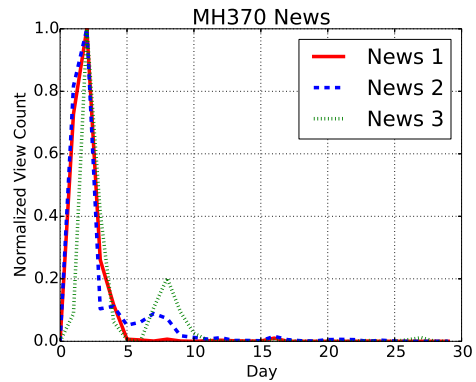}
\caption{Case with constrained eyeball. }
\label{Fig:cssp1}
\end{figure}

The first case is caused by constrained eyeball. The time spent by each user on watching videos is limited.
If there are too many videos telling the same event, these videos will compete with each other for limited
user time. For example, the missing of MH370 is one of the breaking news all over the world in 2014.
The supply of videos about MH370 is overwhelming.  Users with limited watching time cannot cover all videos
about this breaking news. We pick up 3 such videos and plot their view count evolution in Fig.~\ref{Fig:cssp1}.
All curves have very sharp peak, like the news about MH370 plotted in Fig.~\ref{Fig:csnews}.
The difference lies in the total views. These three videos are very unpopular with total views no more than $4000$
during our measured time window. We believe these videos compete with each other such that their total attractiveness
are constrained.

\begin{figure}[ht]
\centering
\includegraphics[width=0.38\textwidth]{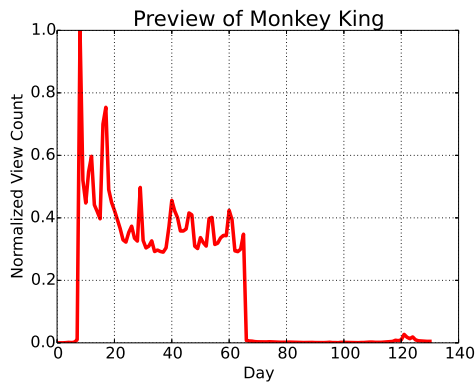}
\caption{Case with interest shift. }
\label{Fig:cssp2}
\end{figure}

The second case is the sudden change of video attractiveness.
In our model, the attractiveness value $q$ is a constant. However, this does not model all cases in real world.
``The Monkey King'' is a very popular movie in 2014 directed
by Hong Kong director Soi Cheang Pou-Soi. It costs $0.6$ billion HK dollar and involves many popular stars. Preview of this movie is very popular, played about $3$ million times
within the first month. View count evolution of this preview is plotted in Fig.~\ref{Fig:cssp2}.
Different from the normal case, there is a sharp view count drop on the 63rd day. Actually, it is
the day that the full movie is uploaded. User attention moves from the preview to the full version video such
that the preview attractiveness diminishes.

\section{Related Work}\label{Sec:related}

Large-scale video delivery over Internet has attracted great attention in both industry and academia during the past few years.
By assuming stationary video popularity, \cite{rossINFOCOM09, zhouINFOCOM12} simplified the performance analysis in P2P VoD systems.
Tu et al. \cite{caoTSMCA13} studied an efficient data scheduling scheme for P2P IPTV system, especially targeting those storage-limited devices. In \cite{massoulieTON13}, Tan et al. studied the content replication scheme in a P2P system consisting of ordinary user devices and set top boxes. Wang et al. \cite{wuTNSM13} proposed a P2P scheme for social network based video service. Niu et al. \cite{libaochunINFOCOM12} studied a cloud bandwidth auto-scaling scheme to meet various user demands while at the same time minimize the bandwidth cost. However, all these works adopted over simplified popularity evolution
models. Our popularity evolution model can be used by these works to refine their schemes.

Many previous works studied video popularity and popularity evolution based on measurement. For example, Cha et al. \cite{chaTON09} studied the popularity distribution and evolution for UGC content by collecting data traces from two large UGC video systems. Li et al. \cite{li2012watching} measured user behavior in the PPTV mobile video platform and developed a practical CDN replication scheme accordingly. In \cite{chenICCCN14}, Chen et al. proposed a lifetime model of online video popularity evolution. \cite{pplive13} incorporated video popularity decaying effect into P2P replication scheme so as to improve delivery performance. However, without theoretical analysis, these works cannot bring out the factors affecting popularity evolution.

Epidemic model originally used to study the spreading of disease\cite{epidemic} in human society has been broadly used to analyze information diffusion in complex networks, e.g. \cite{epidemicPHY01}. Researchers have proposed several different epidemic models to study the information spreading in online social network over the past few years. For example, \cite{leskovecICDM2010} proposed a linear influence model, which does not require the knowledge of the underlying social network and \cite{moonINFOCOM13} used a branching process model to study the message spreading in a microblog service. In \cite{massoulieINFOCOM13}, Jiang et al. studied how information propagates in a social network with limited user attention. Inspired by these works, we use epidemic model to study the spreading of
video information by direct recommendation and word-of-mouth recommendation.

\section{Conclusion}\label{Sec:conclusion}

In this paper, we analyze video popularity and how it changes over time based on data collected from a large-scaled online video content provider in China. Based on this analysis, we come up with a stochastic fluid model to capture the likely factors driving online video popularity dynamics. Besides the video's intrinsic attractiveness (interest level), we believe the fashion the video is recommended (made known) to the viewers, and the viewers¡¯ reaction rate (limited by finite attention span) will together play a major role. Yet other factors could be easily added by extending our model. We validate our model through case studies of common video popularity evolution patterns from our measurements. Our model is a first step towards not only explaining various video popularity dynamics, but also explaining why they behave so. In our validation study, we also bring out some cases that cannot be represented well by our current model, because of some assumptions we made to keep the model simple. For example, the user reaction rate and video attractiveness are assumed to be independent and constant, whereas they may depend on detailed marketing strategy, and may depend on marketing of and the amount of other videos. We will consider these angles in our future work.

\bibliographystyle{IEEEtran}
\bibliography{ref}

\section*{Appendix}
\noindent{\bf Derivation of Eq.~\ref{EQ:xt}, Eq.~\ref{EQ:yt} and Eq.~\ref{EQ:st}:}

For simplicity, in derivation we let  $x$, $y$, $s$ represent $x(t)$, $y(t)$, $s(t)$, respectively.
Since $\frac{dy}{dx} = \frac{1-q}{q}$, we have $y=\frac{1-q}{q}x$. Together with $x+y+s =N$,  we have $s=N-\frac{x}{q}$.
By substituting it back to Eq.~\ref{EQ:dxt}  we have
\begin{eqnarray*}
  &&  \quad \quad \frac{dx}{dt} = \alpha qN + (\beta qN-\alpha)x - \beta x^2 \\
   &\Rightarrow& \quad \int \frac{dx}{\alpha qN + (\beta qN-\alpha)x - \beta x^2} = \int dt \\
  &\Rightarrow& \quad \frac{1}{\alpha + \beta qN} \int \Big( \frac{1}{x+\frac{\alpha}{\beta}} - \frac{1}{x-qN} \Big)dx = \int dt \\
  &\Rightarrow& \quad \frac{1}{\alpha + \beta qN} \ln \Big|\frac{x+\frac{\alpha}{\beta}}{x-qN}\Big| = t + C \\
  &\Rightarrow& \quad \Big|\frac{x+\frac{\alpha}{\beta}}{x-qN}\Big| = e^{(\alpha + \beta qN)(t+C)} = g(t) \\
\end{eqnarray*}
From Lemma~\ref{LEM:Limitxys}, $x(t) \leq Nq$. We have
\begin{eqnarray*}
 &&  \ \Big|\frac{x+\frac{\alpha}{\beta}}{x-qN}\Big| = \frac{x+\frac{\alpha}{\beta}}{qN-x}= e^{(\alpha + \beta qN)(t+C)} = g(t) \\
  &\Rightarrow& \quad x = \frac{qNg(t)-\frac{\alpha}{\beta}}{g(t)+1}
\end{eqnarray*}
Finally, by letting $x(0) = 0$ we obtain $C = \frac{1}{\alpha+\beta qN} \ln \frac{\alpha}{\beta q N}$ and
$g(t) = e^{(\alpha+\beta q N)t+ \ln \frac{\alpha}{\beta q N}}$. Since $\frac{dy}{dx}  = \frac{1-q}{q}$ and $s= N-\frac{x}{q}$,
we can derive $y(t)$ and $s(t)$ correspondingly.
\done
\noindent{\bf Proof of Proposition~\ref{Prop:dxtcurve1}:}

If $\alpha>\beta qN$, $g(t) = \frac{\alpha}{\beta qN}e^{(\alpha+\beta qN)t}>1$ with $t\geq 0$.
Thus, $x''(t)$ is always less than $0$ with $t>0$, $x'(t) $ is a monotonically decreasing function with time $t$.

If $\beta qN<\alpha< (2+\sqrt{3})\beta qN$, we have $t_1>0$ and $t_2<0$. In $[0,t_1]$, $x^{(3)}\leq 0$ and $x''<0$,
thus $x'(t)$ decreases as a concave function. In $[t_1, +\infty)$, $x^{(3)}\geq 0$ and $x''<0$, $x'(t)$ decreases as a convex function.

If $\alpha> (2+\sqrt{3})\beta qN$, we have $t_1<0$ and $t_2<0$. Thus, $x^{(3)}\geq 0$ with $t>0$.
$x'(t)$ decreases as a convex function.

\done
\noindent{\bf Proof of Proposition~\ref{Prop:dxtcurve2}:}

If $\alpha<\beta qN$, $t'>0$.  In $[0,t']$ $x''(t)>0$; while in $[t', +\infty)$ $x''(t)<0$.
Thus, $x'(t) $ is monotonically increasing in $[0,t']$ and monotonically decreasing in $[t',+\infty)$.

If $\alpha< (2-\sqrt{3})\beta qN$, we have $0<t_2<t'<t_1$. In $[0,t_2)$, $x^{(3)} >0$ and $x''>0$,
$x'(t)$ increases as a convex function. In $(t_2, t')$, $x^{(3)}<0$ and $x''>0$,
$x'(t)$ increases as a concave function. In $(t', t_1)$, $x^{(3)}<0$ and $x''<0$,
$x'(t) $ decreases as a concave function. In $(t_1, +\infty)$, $x^{(3)}>0$ and $x''<0$,
$x'(t) $ decreases as a convex function.

If $ (2-\sqrt{3})\beta qN<\alpha<\beta qN$, we have $t_2<0<t'<t_1$. In $(0, t')$, $x^{(3)}<0$ and $x''>0$,
$x'(t)$ increases as a concave function. In $(t', t_1)$, $x^{(3)}<0$ and $x''<0$,
$x'(t) $ decreases as a concave function. In $(t_1, +\infty)$, $x^{(3)}>0$ and $x''<0$,
$x'(t) $ decreases as a convex function.

\done
\noindent{\bf Derivation of Eq.~\ref{EQ:zt}:}

The time period $[0,t]$ is split into $K$ time intervals, each duration is $\Delta$.
During time interval $i$, the increase of $x$ is about equal to $x(i\Delta)\Delta$.
These users are also members in $\mathcal{Z}(i\Delta)$. Then,
\begin{eqnarray*}
   z(K\Delta) & = & \sum_{i=1}^{K}x'(i\Delta)\Delta e^{-(t-i\Delta)\gamma}.
\end{eqnarray*}
Letting $\Delta$ approach $0$ and $K\Delta =t$, we get
\begin{eqnarray*}
   z(t) & = & \lim_{\Delta\rightarrow 0} z(K\Delta)= \lim_{\Delta\rightarrow 0} \sum_{i=1}^{K}x'(i\Delta)\Delta e^{-(t-i\Delta)\gamma}\\
   & = &e^{-\gamma t} \int_{t_0=0}^{t_0=t} x'(t_0)e^{\gamma t_0}dt_0.
\end{eqnarray*}

\done
\noindent{\bf Proof of Proposition~\ref{Prop:dwt}:}

The time is split into time slots. Without a little bit abusing notations, we
let $t$ be a time slot. The numbers of users in $\mathcal{Z}_t$ and $\mathcal{X}_t$ at the
beginning of each time slot are $z(t)$ and $x(t)$.
At the beginning of each time slot, a user makes a decision to view the video with probability $\gamma$.
$dx(t)$ and $dz(t)$ are the change of user population of $\mathcal{X}_t$ and $\mathcal{Z}_t$ respectively.
Then, $dz(t) = -\gamma z(t) + dx(t)$.

If $dx(t)$ is an increasing sequence and $z(t)$ is also increasing, then $dx(t)<dx(t+1)$ and $\gamma z(t)<dx(t)$.
$z(t+1) = z(t)(1-\gamma)+dx(t)$ is the number of users who can make view decision at the beginning of time slot $t+1$.
Since $z(t)<\frac{dx(t)}{\gamma}$, we have $(z(t)(1-\gamma)+dx(t))\gamma<dx(t)<dx(t+1)$ and thus $dz(t+1) = -(z(t)(1-\gamma)+dx(t))\gamma +dx(t+1)>0$.
$z(t)$ is still an increasing sequence at time slot $t+1$.

If $dx(t)$ is a decreasing sequence and $z(t)$ is also decreasing, then $dx(t)>dx(t+1)$ and  $\gamma z(t)>dx(t)$.
Again, $z(t)(1-\gamma)+dx(t)$ is the number of users who can make view decision at the beginning of time slot $t+1$.
This time $z(t)>\frac{dx(t)}{\gamma}$ such that $(z(t)(1-\gamma)+dx(t))\gamma>dx(t)>dx(t+1)$ and thus $dz(t+1) = -(z(t)(1-\gamma)+dx(t))\gamma +dx(t+1)<0$.
$z(t)$ is an decreasing sequence at time slot $t+1$.

Now, let us go back to recall Proposition~\ref{Prop:dxtcurve1} and Proposition~\ref{Prop:dxtcurve2}, there are two cases:
$dx(t)$ is a monotonically decreasing function; $dx(t)$ is a monotonically increasing function in $[0,t']$ and monotonically
decreasing function in $[t', +\infty)$.

For the first case, at the beginning $z(t)\gamma< dx(t)$ since users in $\mathcal{Z}_t$ come from $\mathcal{X}_t$.
With time $t$ goes on, $z(t)$ increases and  $dx(t)$ decreases. Once $z(t)$ begins to decrease, together with the fact that $dx(t)$
is a decreasing sequence, $z(t)$ will keep decreasing in the future.

For the second case, $dx(t)$ keeps increasing before $t'$. Since users in $\mathcal{Z}_t$ come from $\mathcal{X}_t$,
$dz(t)$ begins to increase from $t=0$. Based on above argument, $z(t)$ will keeps increasing with time $t$ until $t'$.
Then, $dx(t)$ begins to decrease and $z(t)$ may still increase. Once $z(t)$ begins to decrease, $z(t)$ will keep decreasing in the future.

In summary, $z(t)$ has a unique peak occurring no earlier than the time when $dx(t)$ achieves peak value.
Since $\frac{dw(t)}{\gamma z(t)}$, the trace of $\frac{dw(t)}{dt}$ is similar to $z(t)$ except a scalar $\gamma$.
\done

\end{document}